\documentclass[aps,prl,preprint]{revtex4}
\usepackage{CJK}

\usepackage{mathtext}
\usepackage{graphicx}
\def\ImageFileParEps#1#2{\centering\includegraphics[#1]{#2.eps}}

\def\vec#1{{\bf #1}}
\def\Av#1{\left<#1\right>}

\def\&{$\!\!$&$\!\!$}

\begin{document}

\begin{CJK*}{GB}{} 
\title{Hot phonon effects and suppressed Auger recombination on 3 $\mu$m  room temperature lasing in HgTe-based multiple quantum well diodes}

\author{A.A.~Afonenko}
\author{D.V.~Ushakov}
\affiliation{Belarusian State University, 220030 Minsk, Belarus}
\author{A.A.~Dubinov}
\author{V.Ya.~Aleshkin}
\author{S.V.~Morozov}
\author{V.I.~Gavrilenko}
\affiliation{Institute for Physics of Microstructures, Russian Academy of Sciences, 603950 Nizhny Novgorod, Russia}
\affiliation{Lobachevsky State University, 603950 Nizhny Novgorod, Russia}

\maketitle
\end{CJK*}

\sloppy



\begin{abstract} 
We propose an electrically pumped laser diode based on multiple HgTe quantum wells with band structure engineered for Auger recombination suppression.
A model for accounting for hot phonons is developed for calculating the nonequilibrium temperature of electrons and holes. Using a comprehensive model accounting for carrier drift and diffusion, Auger recombination and hot-phonon effects, we predict of lasing at $\lambda \sim 3$ $\mu$m at room temperature in 2.2 nm HgTe/Cd$_{0.85}$Hg$_{0.15}$Te  quantum well heterostructure.
  The output power in the pulse can reach up to 600 mW for 100 nanosecond-duration pulses.
\end{abstract}

\section{\label{sec:introduction}Introduction} 

The radiation sources of wavelength around 3 $\mu$m in the atmospheric transparency window is of extreme importance for gas sensing applications, environmental control, and medical diagnostics due to fundamental absorption lines of different substances, for example, paramylon \cite{2021/Zhong/SpectrochActaA/Characteristic}, n-Butanol, Ethene, Ethylene oxide \cite{2004/Sharpe/ApplSpectr/Gas-phase}. Laser sources with this wavelength are also required for lidar systems \cite{2019/Romanovskii/OptLasTech/Development} and free space optical communication systems \cite{2010/Soibel/IEEEPhotTechLet/Midinfrared}. Laser action with wavelength around 3 $\mu$m was realized in semiconductor systems based on III--V (GaSb, InAs), IV--VI (PbS, PbSe), and II--VI compounds \cite{2014/Nasim/OptLasTech/Diode,%
2017/Jung/JOpt/Next-generation}. However, they are either multilayer interband cascade structures, or four and five component solid solutions, which require the highest level of production and limit their distribution. In this regard, the sources based on CdHgTe with HgTe quantum wells (QWs) are very compelling. HgCdTe technology is well-developed and is used for the fabrication of IR detectors \cite{2005/Rogalski/RepProgPhys/HgCdTe,%
2011/Singh/OptLasTech/HgCdTe}. Earlier, lasers based on bulk CdHgTe layers (with optical \cite{1966/Melngailis/APL/Spontaneous} and electrical pumping \cite{1993/Arias/SemSciTech/HgCdTe,%
1993/Million/JCrystGrowth/HgCdTe}) and based on wide CdHgTe QWs with a significant cadmium content  $x=$ 0.5--0.34 (only with optical pumping \cite{1994/Le/APL/Hight,%
1995/Bonnet/JAP/Optical,%
1996/Bonnet/JCrystGrowth/Emission}) were created. It is found that lasing is suppressed due to an higly increasing Auger coefficient from $4.9\cdot 10^{-28}$ cm$^6$/s ($x=0.5$) to $9\cdot 10^{-26}$ cm$^6$/s ($x=0.34$)  at 100 K with a decrease in the band gap of active layers~\cite{1996/Bonnet/JCrystGrowth/Emission} in 1.9--3.5 $\mu$m wavelength range. 

However, due the progress in molecular beam epitaxy the active region of HgCdTe-based emitters can be formed not only by bulk layers but by also by arrays of the binary HgTe QWs. In such quantum well heterostructures it is easier to achieve population inversion as suggested earlier \cite{1998/Vurgaftman/OE/Hight}. As shown in recent works \cite{2018/Alymov/PRB/Auger,%
2019/Aleshkin/JPhysCond/Threshold,%
2020/Alymov/ACSPhot/Fundamental}
when using HgTe active layers it is possible to suppress Auger recombination due to Dirac-like energy dispersion. As a result, interest in this system of materials was revived, and with optical pumping it was possible to observe stimulated emission at wavelengths of 3.7 \cite{Kudryavtsev_TempLimitations}, 2.8 \cite{2018/Fadeev/OptExp/Stimulated}, and 
2.45~$\mu$m~\cite{2021/Fadeev/OptEng/MidIR} at temperatures of 240, 250, and 300 K, respectively.

The main factors that limit the temperature characteristics of mid-infrared (IR) semiconductor lasers are Auger recombination on the QWs, non-equilibrium carrier heating and Drude absorption of the free carriers. Although yet there is no experimental demonstration of injection lasers based on heterostructures with HgTe QWs, there are several theoretical proposals of diode laser aimed for mid- \cite{1998/Vurgaftman/OE/Hight} and far-IR \cite{2021/Afonenko/JPhysD_ApplPhys/Feasibility} operation as well as THz quantum cascade laser \cite{2020/Ushakov/OE/HgCdTe} based on HgCdTe. In this work given the experimental results demonstrating stimulated emission under optical pumping \cite{Kudryavtsev_TempLimitations,%
2018/Fadeev/OptExp/Stimulated,%
2021/Fadeev/OptEng/MidIR} and suppression of Auger recombination in HgTe QWs \cite{2018/Alymov/PRB/Auger,%
2020/Utochkin/FTP/Continuous-Wave} we investigate the feasibility of HgCdTe-based injection lasers. Considered laser design has active region consisting of several 2.2-nm thick HgTe QWs separated by Cd$_x$Hg$_{1-x}$Te barriers. The QW width is selected to achieve the emission wavelength of 3 $\mu$m. 
Here using a complex numerical model that takes into account the drift and diffusion of carriers in the barrier layers, carrier capture in the quantum wells, radiative and non-radiative recombination and heating of the active region we show the possibility of generation at room temperatures. We develop a model which accounts for hot phonons and show the essential influence of non-equilibrium heating of charge carriers on the output characteristics of lasers, especially for structures with deep QWs and low optical phonon energy.

\section{Structure design}

We consider the following multiple quantum well laser structure grown on a (013)-oriented $n$-GaAs substrate: 100~$\mu$m $n$-GaAs substrate / 5~$\mu$m $n$-CdTe cladding / 600 nm $n$-Cd$_x$Hg$_{1-x}$Te waveguide ($x$ varies between 0.6 and 1) /active region (five--ten 2.2 nm HgTe QWs interleaved with 10 nm Cd$_{0.85}$Hg$_{0.15}$Te barriers) / 70~nm $p$-Cd$_x$Hg$_{1-x}$Te waveguide ($x$ varies between 0.85 and 1.0) / 3 $\mu$m $p$-CdTe cladding.
The calculated parameters of each heterostructure layer are collected in Table~\ref{tab:parameters}.

\begin{table}[!t]
\caption{\label{tab:parameters}
Parameters of heterostructure layers: layer number (\#), thickness ($d$), composition, doping, electron ($\mu_n$) and hole ($\mu_p$) mobilities, bandgap ($E_g$)  and real part of the refractive index ($n'$) for the $\sim 3\,\mu$m wavelength. All the quantities are for 3D materials and temperature 300 K. In the graded-composition waveguide layers, the parameters vary linearly between their values in adjacent layers.}

\begin{tabular}{*{8}c}
\hline
\hline
\# \& $d$  \& Composition \& Doping  \& $\mu_n$  \& $\mu_p$ \& $E_g$ \&   $n'$\cr
 \&  (nm) \&  \&  (cm$^{-3}$) \& ($\frac{\rm cm^2}{\rm V \cdot s}$) \& ($\frac{\rm cm^2}{\rm V \cdot s}$) \&  (eV)  \& \cr
 \hline
\multicolumn{8}{c}{Substrate}\\
1 \&  \& GaAs \& $n: 10^{18}$ \& 2361 \& 134 \& 1.42 \&  3.31 \cr \hline
\multicolumn{8}{c}{$n$-cladding layer}\\
2 \& 5000 \& CdTe \& $n: 10^{17}$ \& 111 \& 36 \& 1.49 \&   2.71\cr \hline
\multicolumn{8}{c}{$n$-waveguide layers}\\
\hline
3 \& 50 \& CdTe \& $n: 5\cdot 10^{17}$ \& 40 \& 16 \& 1.49 \& 2.7\cr 
4 \& 50 \& graded \cr 
5 \& 400 \& Cd$_{0.6}$Hg$_{0.4}$Te \& $n: 10^{17}$ \& 197 \& 36 \& 0.71   \& 2.86\cr
6 \& 50 \& graded \cr 
7 \& 30 \& Cd$_{0.9}$Hg$_{0.1}$Te \& $n:10^{18}$ \& 30 \& 11 \& 1.26  \& 2.74\cr
8 \& 10 \& graded \cr 
9 \& 10 \& Cd$_{0.85}$Hg$_{0.15}$Te \& $n:10^{16}$ \& 523 \& 73 \& 1.15  \& 2.78\cr
\hline
\multicolumn{8}{c}{active region}\\
10 \& 2.2 \& QW: HgTe \& 0 \& 523 \& 73 \& -0.14 \& 5.14\cr
11 \& 10 \& Cd$_{0.85}$Hg$_{0.15}$Te \& $n:10^{16}$ \& 523 \& 73 \& 1.15  \& 2.78\cr
\multicolumn{8}{c}{ + (5$\div$9) + 1/2 periods (2 layers in period)}\\
\hline
\multicolumn{8}{c}{$p$-waveguide layers}\\
12 \& 10 \& Cd$_{0.85}$Hg$_{0.15}$Te \& $n:10^{16}$ \& 523 \& 73 \& 1.15 \& 2.78\cr
13 \& 10 \& graded\cr
14 \& 20 \& CdTe \& $p: 5\cdot 10^{17}$ \& 40 \& 16 \& 1.49  \& 2.71\cr 
15 \& 30 \& CdTe \& $p: 1\cdot 10^{16}$ \& 414 \& 73 \& 1.49  \& 2.71\cr 
\hline
\multicolumn{8}{c}{$p$-cladding layer}\\
16 \& 3000 \& CdTe \& $p: 4\cdot 10^{16}$ \& 194 \& 51 \& 1.49  \& 2.71\cr 
\hline
\multicolumn{8}{c}{Au contact  layer}\\
\hline
\hline
\end{tabular}

\end{table}

Band structure parameters (bandgap, effective masses) of ternary and binary compounds were obtained from experimental and theoretical data using the approach of Refs.~\cite{2005/Novik/PhysRevB/Band,2016/ALESHKIN/PhysB/Effect}.
Electron and hole mobilities were interpolated from the experimental data of \cite{1972/Scott/JAP/Electron,%
2018/Nagaoka/APL/High}. Infrared refractive index and absorption coefficient of the waveguide were calculated taking into account Drude and lattice absorption~\cite{1983/Mroczkowski/JAP/Optical,1990/Laurenti/JAP/Temperature,1974/Grynberg/PhysRevB/Dielectric}. Internal optical losses and refractive index were calculated within the multioscillator Lorentz-Drude model including both phonon and free-carrier contributions~\cite{1983/Mroczkowski/JAP/Optical,1990/Laurenti/JAP/Temperature,1974/Grynberg/PhysRevB/Dielectric}.
Calculated absorbtion cross sections of the free electrons and holes were $1.7\cdot 10^{-17}$  and $1.2\cdot 10^{-16}$ cm$^{2}$ respectively.  Quantum well depths for electrons and holes were calculated according to Refs.~\cite{2005/Novik/PhysRevB/Band,1989/Van_de_Walle/PhysRevB/Band}. 

\begin{figure}
\includegraphics[width=0.5\textwidth]{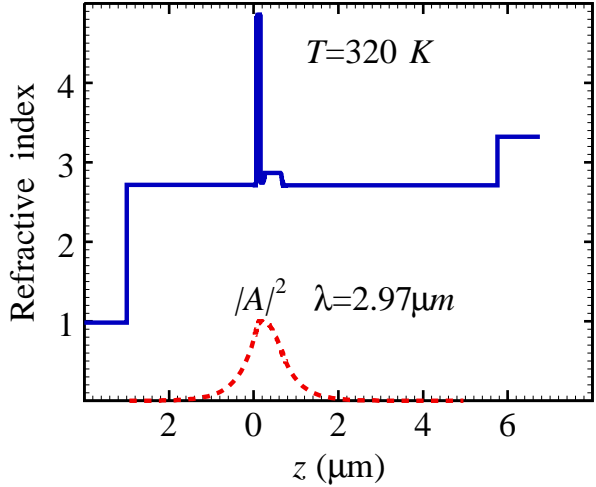}%
\caption{ Spatial distribution of the refractive index across the simulated heterostructure at $\lambda = 2.97$ $\mu$m and the squared electric field of the ground TE mode. The total optical confinement factor of 8 quantum wells is $\Gamma = 0.014$, the effective refractive index is $n_{\rm eff}$ = 2.78. 
}\label{fig:optical_properties}
\end{figure}

Refractive index was interpolated from the Ref~\cite{1987/Kucera/PSSa/Dispersion} for the Cd content in the range 0.18--1.0. During the interpolation the background permittivity caused by the transitions involving L and X valleys was found analogously to Ref \cite{1988/Pikhtin/FTP/Refraction}. Resonant input of interband transitions for bulk semiconductor was calculated according to \cite{1987/Adachi/PRB/Model}.  At the same time for HgTe the experimental data is scarce. We did not find the direct measurements of the refractive index of HgTe in mid-IR range. The available experimental works present the measurements of refractive index in the range of phonon absorption  19--20 meV and photon energies above 1.5 eV \cite{1998/Palik/Handbook}. Thus in order to calculate the refractive index of HgTe QWs we used combination of the background permittivity extrapolated to $x=0$ from the results obtained in Ref \cite{1987/Kucera/PSSa/Dispersion} and resonance term related to step-function density of states caused by dimensional quantization. The obtained value of refractive index for HgTe QW at $\lambda=2.97$~$\mu$m is equal to 5.14 (see Table \ref{tab:parameters}). It agrees with the value of $\sim$4.6 obtained in Ref.~\cite{1985/Jones/APL/Infrared}, for HgTe(4 nm)/CdTe(2 nm) superlattice at the wavelength of its absorption edge equal to 6.7 $\mu$m. This refractive index is a weighted average composed of 2/3 of the refractive index of HgTe 5.6 and 1/3 of the refractive index of CdTe 2.71. Note that within the calculations of electromagnetic modes we also take into account carrier induced refractive index change, which under conditions of injection decreases the refractive index to 4.8 (Fig. \ref{fig:optical_properties}).

In order to reduce Auger recombination in the well, the mole composition of the barrier layers was chosen to be $\sim 0.85$, which corresponds to a refractive index of 2.78, which is slightly higher than the refractive index of CdTe 2.71. Because of the small change of refractive indices and small thickness of the core, the waveguide effect turned out to be very weak. Therefore, additional waveguide layers around the active region are required. In our calculations we used Cd$_{0.6}$Hg$_{0.4}$T with 400 nm thickness and refractive index of 2.86 (see Fig.~\ref{fig:optical_properties}). Total calculated optical confinement factor is $\Gamma \approx 0.0018$  per QW. Suggested design also includes high Cd$_{0.9}$Hg$_{0.1}$Te barrier placed between the active and waveguide layers to prevent parasitic interband recombination in the waveguide layer due to hole injection into it.

The reflection coefficients of the output facet for the TE$_0$ mode was chosen equal to Fresnel reflection $r = 0.2$. The opposite facet was considered with a high reflection coating with a reflection coefficient $r = 0.9$.

\section{Theoretical model}

This structure was simulated using a distributed drift-diffusion model based on one-dimensional (1D) Poisson's equation and continuity equations for electrons and holes with taking into account the  carrier capture and  escape processes~\cite{2014/Afonenko/FTP/Current}. Both radiative and Auger recombination were included in the model.

The band structure parameters were obtained from the eight-band $\vec{k} \cdot \vec{p}$ method~\cite{2005/Novik/PhysRevB/Band,2016/ALESHKIN/PhysB/Effect}. Quantum well depths for electrons and holes were calculated according to Refs.~\cite{2005/Novik/PhysRevB/Band,1989/Van_de_Walle/PhysRevB/Band}. Carrier mobilities were interpolated from the experimental data of \cite{1972/Scott/JAP/Electron}.

\begin{figure}
\includegraphics[width=0.49\textwidth]{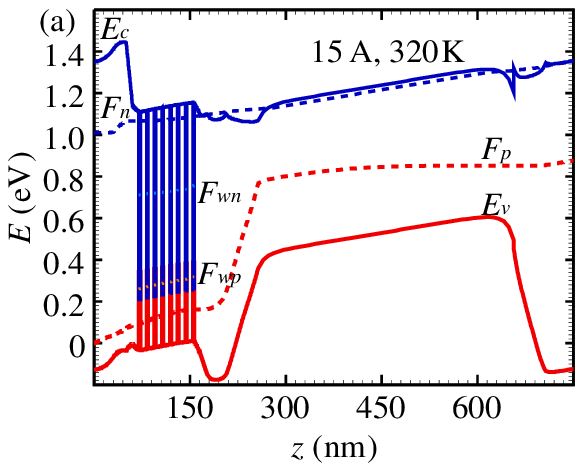}%
\includegraphics[width=0.49\textwidth]{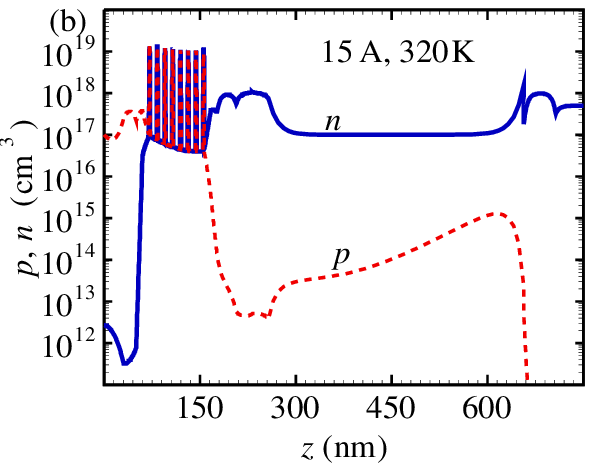}%
\caption{
(a) Band diagram and (b) distribution of carrier densities inside the 8~QW active region of the simulated heterostructure, calculated at
320 K, 1.357~V bias, 15 A drive current  (above the lasing threshold: photon density $S^{(2D)}=1.25\cdot 10^{11}$~cm$^{-2}$). $F_{wn}$, $F_{wp}$
($F_n$, $F_p$) are the quasi-Fermi levels of localized (delocalized) carriers.
}\label{fig:bands_and_concentrations}
\end{figure}

An example of calculated band diagrams of structure with 8 QW and carrier distributions are shown in Fig.~\ref{fig:bands_and_concentrations}. In order to reduce optical losses, doping of most of the layers was selected to be below $10^{17}$ cm$^{-3}$. This results in a large ohmic resistance and a noticeable decline of the of quasi Fermi levels in the emitters. Gradient layers reduce the space charge region, while introduction of a wide-gap Cd$_{0.9}$Hg$_{0.1}$Te blocking layer creates a barrier for the holes, preventing excitation of the waveguide.

Distribution of the photon density across the resonator was found from the Bouguer--Lambert--Beer law. 

Heating effects were taken into account by solving the 1D heat equation in the direction perpendicular to the heterostructure layers (we consider pulsed operation, when in-layer heat transfer during a pulse is negligible). We also included the temperature dependence of the bandgap in our model.

\section{Auger recombination}\label{sec:Auger}

In our calculations of the interband recombination we included both the radiative and Auger processes 
\cite{2021/Afonenko/JPhysD_ApplPhys/Feasibility,%
1959/Beattie/ProcRSoc/Auger,%
1991/Jiang/JAP/Carrier,%
1998/Polkovnikov/PRB/Auger}. 
In this work, the coefficient of Auger recombination was calculated numerically. For a process involving two electrons and one hole, the coefficient $C_{nnp}$ was found from the expression:
\begin{equation}\label{eq:Auger_coefficient}
\begin{array}{ll}
C_{nnp}n^2p=&\displaystyle\frac{2\pi}{\hbar}\int\!\!\!\int\!\!\!\int
\left<\left|V_{1234}^2\right|\right>
f_c\left(E_1\right)f_c\left(E_2\right)f_v\left(E_3\right)\times
\\ 
&\displaystyle
\times\left[1-f_c\left(E_4\right)\right]\delta\left(E_1+E_2-E_3-E_4\right)\frac{1}{2}\frac{2d\mathbf{k}_1}{\left(2\pi\right)^2}\frac{2d\mathbf{k}_2}{\left(2\pi\right)^2}\frac{d\mathbf{k}_3}{\left(2\pi\right)^2},
\end{array}
\end{equation}
where $\hbar$ is reduced Planck constant,
$E_{i}$ and $\mathbf{k}_{i}$ -- the energies and the wave vectors of the particles involved in the interaction,
$f_c(E)$ and $f_v(E)$ are Fermi-Dirac functions for electrons and holes,
$n$ and $p$ are the concentrations of electrons and holes,
\begin{equation}
\left<\left|V_{1234}^2\right|\right> 
= \left(\frac{2\pi e^2}{4\pi \epsilon\epsilon_0}\right)^2\frac{1}{4}\sum\limits_{\alpha,\beta,\gamma,\delta}
\left|\frac{V_{4\delta\, 3 \gamma , 1 \alpha\, 2\beta}}{q_{13}}
-\frac{V_{4\delta\, 3 \gamma , 2\beta\, 1 \alpha}}{q_{23}}
\right|^2.
\end{equation}
is averaged square of the matrix element of the Coulomb interaction in a quasi-two-dimensional system, taking into account the exchange interaction,
$q_{ij}=\sqrt{q_{\rm scr}^2 + (\mathbf{k}_i-\mathbf{k}_j)^2}$ ,
$q_{\rm scr}$ is the inverse screening length, 
\begin{equation}\label{eq:Auger_V}
\begin{array}{ll}
\displaystyle
V_{4\delta\, 3 \gamma , 1 \alpha\, 2\beta} = 
& \displaystyle
\int\!\!\!\int
\psi_{c4\delta}^*(z_2)\psi_{v3\gamma}^*(z_1)
 \times
\\ 
& \displaystyle 
\times\exp\left(-q_{13}|z_1-z_2|\right)
\psi_{c1\alpha}(z_1)\psi_{c2\beta}(z_2)
\, dz_1\,dz_2
\end{array}
\end{equation}
-- matrix element of the Coulomb interaction of selected states.
The indices $\alpha$, $\beta$, $\gamma$, $\delta$ number the initial and final states that are doubly degenerate in the electron spin,
$\psi(z)$ -- dependences of wave functions on the coordinate $z$. The wavefunctions and their energies were found by the 8-band $\bf k\cdot p$ method
\cite{2005/Novik/PhysRevB/Band,%
2016/ALESHKIN/PhysB/Effect}.
The inverse screening length was calculated as for 2D system of electrons and holes:
\begin{equation}
q_{\rm scr}  = \frac{e^2}{2\epsilon\epsilon_0}
\left(\frac{\partial n}{\partial F_{wn}}-\frac{\partial p}{\partial F_{wp}}\right).
\end{equation}
Here
$F_n$ and $F_p$ are quasi Fermi levels for electrons and holes.

The coefficients of Auger recombination involving two holes and one electron $C_{npp}$ were calculated by the formulas
(\ref{eq:Auger_coefficient})--(\ref{eq:Auger_V}) with interchange $c\leftrightarrow v$.





For the analyzed quantum well HgTe/Cd$_{0.85}$Hg$_{0.15}$Te 2.2 nm wide for 300 K, the coefficients of Auger recombination for electrons $C_{nnp}=9.2\cdot10^{-16}$~cm $^4$/s and for holes
$C_{npp}=3.8\cdot10^{-15}$~cm$^4$/s.
The level of excitation was taken close to the threshold.

Resonant transitions with a small change in wave vectors ($<q_{\rm scr}$) make a significant contribution to the recombination. In this case, the final high-energy state can be a continuum state. 
For example, for transitions involving two electrons and one hole, this contribution is about is about half of the resulting coefficient, 
while for a nonresonant transition with the participation of localized electrons only in the first subband, the recombination Auger coefficient is $C_{nnp}=4.4\cdot10^{-16} $~cm$^4$/s.
Auger recombination for transitions involving two holes and one electron is determined mainly by resonant transitions, and the corresponding coefficient is 4 times higher than the value for a process involving two electrons and one hole.
This differs from the Auger calculations of recombination without taking into account the continuum states,
when CCCH process with scaterring of two electrons and one heavy hole is the most probable among different types of Auger processes in narrow-gap materials \cite{2020/Alymov/ACSPhot/Fundamental}.
Quantitatively the obtained values of the resulting recombination Auger coefficient $C=C_{nnp}+C_{npp}=4.7\cdot10^{-15}$~cm$^4$/s for $n=p$ is smaller than in calculations \cite{2020/Alymov/ACSPhot/Fundamental}, where only transitions involving localized electrons and holes were taken into account ($C_{nnp}=1.2\cdot10^{-14}$~cm$^4$/s).
However, the exchange interaction, which reduces the recombination Auger coefficient, was not taken into account in \cite{2020/Alymov/ACSPhot/Fundamental}.

Verification of the coefficients of Auger recombination with the experimental data of \cite{Kudryavtsev_TempLimitations} has been carried out. For Cd$_{0.08}$Hg$_{0.92}$Te / Cd$_{0.8}$Hg$_{0.2}$Te quantum well with thickness of 2.7 nm ($\lambda\approx 2.7$~$\mu$m), the measured value at an external temperature of 250 K and an electron temperature of 400 K and the charge carrier concentration of $1.2\cdot10^{12}$~cm$^{-2}$ is $C^{3D}\approx10^{-27}
$~cm$^{6}$/s.
According to our calculations based on formulas (\ref{eq:Auger_coefficient})--(\ref{eq:Auger_V}) for temperature 400~K $C^{3D}=C^{2D}d^2=6.5\cdot10^{-28}$~cm$^{6}$/s.



\def\Nph{N_{\rm ph}}
\def\NphT{N_{\rm ph 0}}
\def\tauph{\tau_{\rm ph}}
\def\qph{q_{\rm ph}}
\def\rhoph{\rho_{\rm ph}}
\def\etaph{\eta_{\rm ph}}
\def\mph{m_{\rm ph}}

\def\hwLO{\hbar\omega_{LO}}
\def\vec#1{{\bf #1}}
\def\Av#1{\left<#1\right>}

\section{Hot phonon effect} \label{sec:HotPhonon}

One of the mechanisms limiting the output power of semiconductor lasers is nonequilibrium heating of electrons in quantum wells. ``Hot'' electrons appear in the active region through the process of their capture from barrier regions to localized levels and subsequent thermalization to the lower levels of the quantum well. Electrons and holes lose energy by emitting optical phonons. The processes of relaxation of the electron and hole temperature to the ``hot'' phonon temperature are faster than the equilibration of the ``hot'' phonon temperature and the lattice temperature. Therefore, it is possible to consider the temperature of electrons and holes equal to the temperature of ``hot'' phonons:

\begin{equation}
\Nph = \frac{1}{\exp\left(\frac{\hwLO}{k_B T}\right)-1}.
\end{equation}
Here $\hwLO$ is the energy of optical phonons.

The balance of ``hot'' phonons $\Nph$ was estimated analogously to \cite{1991/Ridley/RepProgPhys/Hot,%
1992/Ozturk/SemSciTechn/Energy} from the equation:
\begin{equation}\label{HPEqdNphdt}
\frac{d\Nph}{dt} =- \frac{\Nph - \NphT}{\tauph}
 + \etaph \frac{R E_{gb} -  (R_{\rm st}+R_{\rm sp})\hbar\omega}{\rhoph\hwLO}.
\end{equation}
Here $\NphT$ is the equilibrium occupation numbers of phonon modes, 
$\tauph$ is the lifetime of phonon modes, which takes into account the decay of optical phonons into acoustic phonons, scattering processes on interface roughness, etc.
$R$ is the rate of electron (hole) injection in a QW,
$E_{gb}$ is the bandgap energy of the barrier layers, 
$R_{\rm st}$ and $R_{\rm sp}$ are the rates of stimulated and spontaneous recombination, 
$\hbar\omega$ is the energy of emitted photons, 
$\rhoph$ is the density of phonon modes, effectively interacting with electrons and holes.
$\etaph$ is the portion of emitted phonons, which fall into the modes involved in thermalization of the carriers near the bottom of the band.

\begin{figure}
\ImageFileParEps{width=0.45\textwidth}{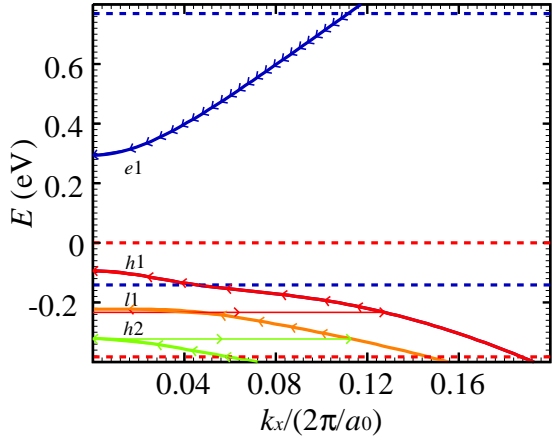}
\caption{
Dispersion of electrons and holes in [013] direction in 2.2~nm thick HgTe/Cd$_{0.85}$Hg$_{0.15}$Te quantum well. Horizontal lines show the energies of the edges of the QW and the barrier layers of the conduction band (blue dashed lines) and the valence band (red dashed lines). Arrows show the transitions of charge carriers associated with emission of optical phonon.}\label{HPFigE(k)}
\end{figure}

For the estimates one can consider an ensemble of the ``hot'' phonons consisting of all the phonons with wavevector smaller than the highest possible value of the wavevector of the emitted phonons:
\begin{equation}
\qph \le \left|\vec{k}_2\right|+\left|\vec{k}_1\right| \approx 2 \left|\vec{k}_2\right|,
\end{equation}
where $\vec{k}_2$ and $\vec{k}_1$ are the wavevectors of the initial and final states of the electron. Approximation is true for the transitions of electrons with the energy $E(k)-E(0)\gg \hwLO$. Scattering with a subband change is less probable than scattering inside one subband. Therefore we assume that first carriers relax inside a subband to its bottom and then they move to the states of the subband with lower energy, as shown in Fig.~\ref{HPFigE(k)}. In this case the density of phonon modes can be obtained after averaging the square of the wavevector of the current carrier with regard to its energy: 
\begin{equation}\label{HPEqRhoPh0}
\rhoph = \frac{\left<k^2\right>_E}{\pi^2}.
\end{equation}
Here, the fraction of emitted phonons falling into localized phonon modes is considered to be $\etaph=1$.

Described method for estimating the nonequilibrium temperature provides underestimated values for overheating, since it does not take into account the uneven distribution of emitted phonons along the wavevectors and the selective interaction of electrons and holes with phonons.

In order to find the distribution of emitted phonons along the wavevectors, we can write the probability of electron transition by emitting optical phonons

\begin{equation}\label{HPEqOmegaQ}
\omega_q=\frac{2\pi}{\hbar}\left|V_q^2\right|\delta\left(E_2-E_1-\hbar\omega_{LO}\right).
\end{equation}
Here $E_{2}$ and $E_{1}$  are the initial and final energies of the electron (hole). The square of the matrix element of electron-phonon interaction with phonon modes, localized near the interfaces, is inversely proportional to the value of the wavevector $q$:
\begin{equation}
V_q^2\sim\frac{1}{q}.
\end{equation}
In balance equation for the population of electrons (holes) it is necessary to take into account the transitions to the adjacent lower level with probability $\omega \sim (\Nph+1)$ and to the adjacent higher level with probability $\omega \sim \Nph$. In this case, the relaxation of an electron or hole to the bottom of the quantum well leads to emission of dozens of optical phonons and the chain of transitions is very long. It can be shown that for an infinite chain of such levels there are two stationary solutions. One solution describes thermodynamic equilibrium when the populations are connected by Boltzmann relation. The second solution is dynamic, with all the populations being the same. In this case, stimulated transitions with phonon emission are compensated by stimulated transitions with phonon absorption and the resulting rate of transitions is determined only by the probability of spontaneous transitions.

We can average (\ref{HPEqOmegaQ}) over all possible directions of the initial wavevector of the electron for a fixed wavevector of the phonon $\vec{q}$ considering $\vec{k}_1=\vec{k}_2-\vec{q}$:
\begin{equation}
p\left(\vec{q}\right)\sim \int\frac{1}{q}\delta\left(\frac{dE_1}{dk_1^2}\left(k_2^2-2k_2q\cos{\phi}+q^2-k_1^2\right)\right)d\phi.
\end{equation}
Here $p\left(\vec{q}\right)$ is the probability density of emitting the phonon with wavevector $\vec{q}$.
After calculating the integral we obtain:
\begin{equation}\label{HPEqPph}
p\left(q\right)=2\pi q p\left(\vec{q}\right)\sim\frac{1}{\sqrt{\left(q_p^2-q^2\right)\left(q^2-q_m^2\right)}}.
\end{equation}
Here $p\left(\vec{q}\right)$ is the probability density to emit a phonon with wavevector equal to $q$,
$q_m=k_2-k_1$, $q_p=k_2+k_1$.
Proportionality factor can be found from the normalization condition by numerically calculating the integral
\begin{equation}
\int\limits_{q_m}^{q_p} p\left(q\right) dq=1.
\end{equation}
Thus, the probability density of emitting the phonon with wavenumber $q$ increases near the range boundaries $[k_m, k_p]$ and decreases in the middle of this range.

In order to find the density of phonon modes effectively interacting with electrons we determine the rate of transitions of electrons near the bottom of a band accompanied by phonon absorption
\begin{equation}
R\sim \int{\omega_q\exp{\left(-\frac{E_1}{k_B T}\right)}\frac{d^2k_1}{\left(2\pi\right)^2}}.
\end{equation}
For simplicity we consider that electron system is non-degenerate and the second term of the integrand takes into account Boltzmann distribution of electrons in the energy levels.
We consider that $E_2=\hbar^2k_2^2/2m_c$, $E_1=\hbar^2k_1^2/2m_c$, $k_2^2=k_1^2+2k_{1x}q+q^2$.
After the integration we can find the weight function, which accounts for relative contribution of different $q$ in electron heating
\begin{equation}
f_c\left(\vec{q}\right) =
\left(\frac{2m_c\hbar\omega_{LO}}{\hbar^2q^2}\right)
\exp\left[-\frac{\hbar^2}{2m_c k_B T}\left(\frac{m_c\hbar\omega_{LO}}{\hbar^2q}-\frac{q}{2}\right)^2\right].
\end{equation}
Similar expression can be obtained for the holes by substituting effective mass of an electron $m_c$ by effective mass of a hole $m_v$. 

Combined weight function which accounts for relative contribution of different $q$ in heating of both electrons and holes and be found by summing the individual taking into account the condition of charge neutrality of the QW (i.e. equal concentration of electrons and holes)
\begin{equation}\label{HPEqFcv}
f_{cv}\left(\vec{q}\right) \sim \frac{f_{c}\left(\vec{q}\right)}{m_c} + \frac{f_{v}\left(\vec{q}\right)}{m_v}.
\end{equation}
Weight function is normalized according to condition $\max{f_{cv}\left(\vec{q}\right)}=1$. With that in mind the density of the phonon modes, which are involved in carrier thermalization near the bottom of the band can be written as:
\begin{equation}\label{HPEqRhoPh}
\rhoph=\int{f_{cv}\left(\vec{q}\right)\frac{d^2q}{\left(2\pi\right)^2}},
\end{equation}
while the proportion of spontaneous phonons which fall into the modes involved in thermalization of the carriers near the bottom of the band is expressed as:
\begin{equation}\label{HPEqEtaPh}
\etaph=\int f_{cv}\left(\vec{q}\right)p\left(q\right)dq.
\end{equation}

Note, that the choice of normalization of the weight function does not affect the final result because the values $\rhoph$ and $\etaph$ are found in balance equation for non-equilibrium phonons (\ref{HPEqdNphdt}) as a fraction. 

The resulting approximate expression for the density of phonon modes due to interaction of one type of the modes with charge carriers reads as:
\begin{equation}
\rhoph= \frac{\mph\sqrt{\hwLO k_B T}}{\sqrt{\pi}\hbar^2}.
\end{equation}
This expression can also be used in a general case by finding the parameter $\mph$ from numerical solution of integral (\ref{HPEqRhoPh}). 

\begin{figure}
\ImageFileParEps{width=0.45\textwidth}{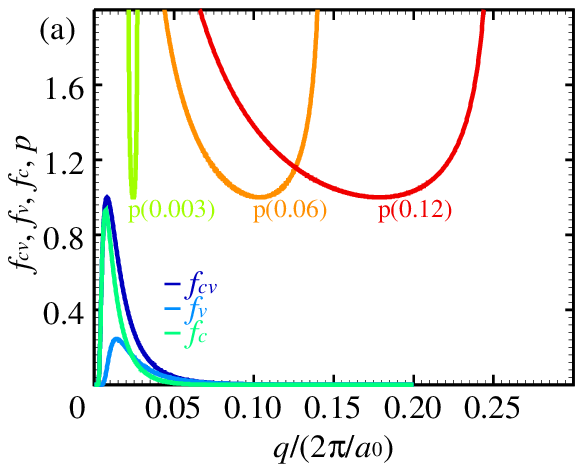}
\ImageFileParEps{width=0.45\textwidth}{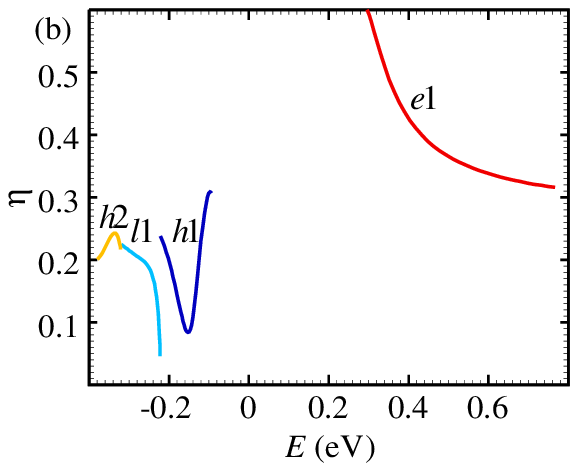}\\

\caption{
(a) Dependence of the weight function for the lowest subbands of the quantum well, its electron and hole components on the wavevector of the phonon at $T=300$~K and probability density (not normalized) of electron emitting the phonon with wavevector of $k=0.003$, 0.06, 0.12 $\times 2\pi/a_0$ ($a_0$ -- lattice constant);
(b) Dependence of the proportion of the emitted phonons involved in carrier thermalizations near the bottom of the band on the energy of charge carriers. Captions $e1$ denotes electron subbands, $h1$, $h2$, $l1$ denotes heavy and light hole subbands respectively.
}\label{HPFigEta(E)}
\end{figure}

As Fig.~\ref{HPFigEta(E)}~a shows, phonons intensely interacting with electrons have smaller values of wavevectors than phonons interacting with holes. This is due to the fact that effective mass of electrons is smaller than effective mass of holes. The dependence of the proportion of emitted phonons involved in carrier thermalization near the bottom of the band has a complex energy dependence determined by carrier dispersion (Fig.~\ref{HPFigEta(E)}~b). This proportion increases near the minima of the lower subbands, but never reaches a value of 1 due to thermal distribution of the carriers. The average value of the proportion of spontaneous phonons involved in carrier thermalization at the bottom of the band is obtained by averaging over the carrier energy. In considered case, $\Av\etaph_E=0.31$ with effective mass of the density of phonon modes $\mph=0.029$ while masses of electrons and holes are 0.050 and 0.19 respectively.

Thus, accounting for non-uniform distribution of emitted phonons along the wavevectors according to equation (\ref{HPEqPph}) and selective by wavevector interaction of electrons and holes with phonons according to equations (\ref{HPEqFcv})--(\ref{HPEqEtaPh}) in the considered case results in 12 times higher heating rate of the non-equilibrium electron-phonon system $\etaph R/\rhoph$ (see equation (\ref{HPEqdNphdt})) compared to the estimates obtained in the approximation of uniform distribution of emitted phonons along the wavevectors from equation (\ref{HPEqRhoPh0}).

\section{\label{sec:results}Results and discussion} 

\begin{figure}
\includegraphics[width=0.5\textwidth]{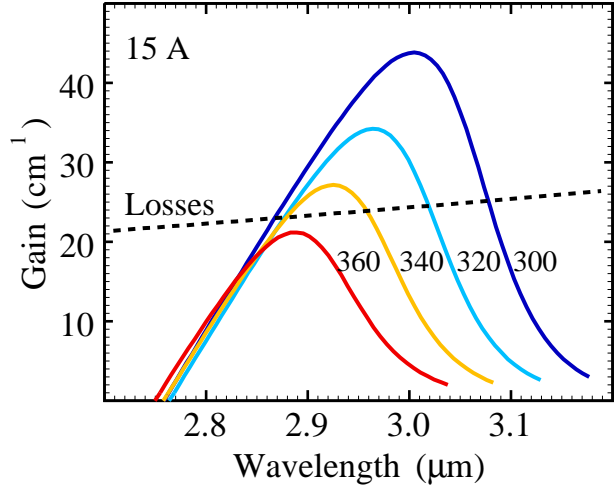}%
\caption{Gain spectra (unsaturated: photon density $S^{(2D)}=0$) for the TE$_0$ mode at lattice temperatures $T = $300--360~K. The corresponding drive current density is $J = 15$ A. The estimated total losses are also shown with dashed line.
}\label{fig:gain}
\end{figure}

One way to obtain the necessary amplification at moderate rate of Auger recombination is to use multiple QW instead of one. According to our calculations, $N_{\rm QW} = $6--10 is the optimal number of QWs, which allow to decrease Auger recombination rate with no significant increase in free carrier absorption and threshold current.
The optimized current and optical confinement result in modal gain exceeding the total optical loss (estimated as 22--24 cm$^{-1}$ at $\lambda = 3$~$\mu$m in the simulated structure) up to $\sim 350$ K (Fig.~\ref{fig:gain}), thus allowing laser operation  above room temperature. The gain maximum is shifted to shorter wavelengths with increasing temperature due to the temperature dependence of the bandgap.

\begin{figure}
\includegraphics[width=0.5\textwidth]{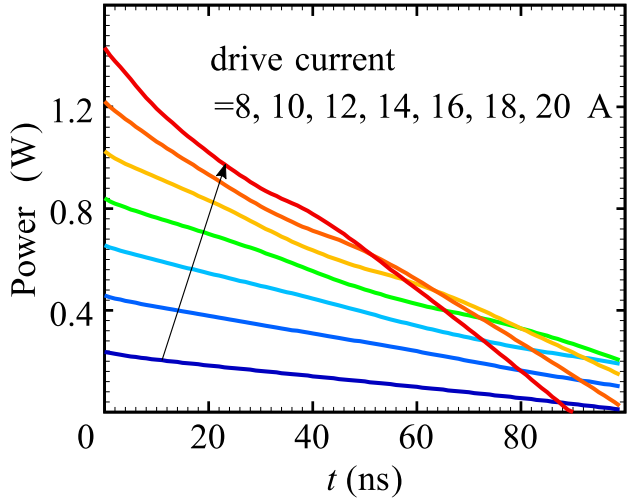}%
\caption{Time evolution of output power of the active region during a 100 ns pulse at different drive currents. The heat sink temperature is 300 K. 
}\label{fig:P_t_I}
\end{figure}

\begin{figure}
\includegraphics[width=0.5\textwidth]{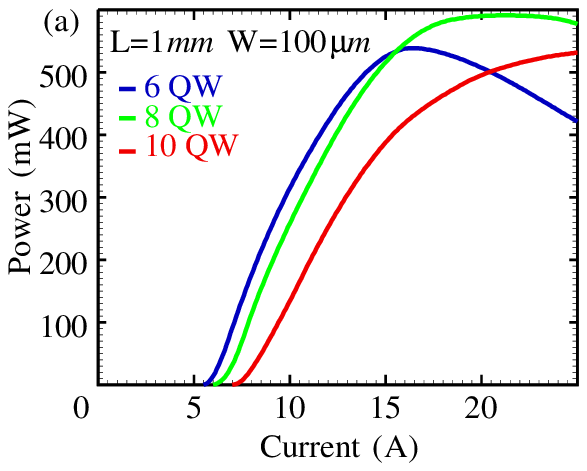}%
\includegraphics[width=0.5\textwidth]{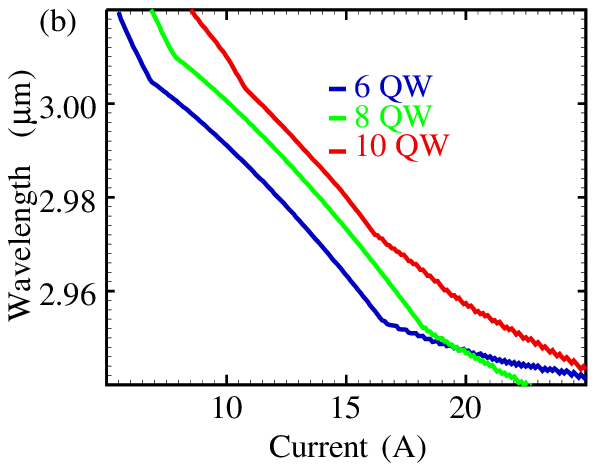}\\
\includegraphics[width=0.5\textwidth]{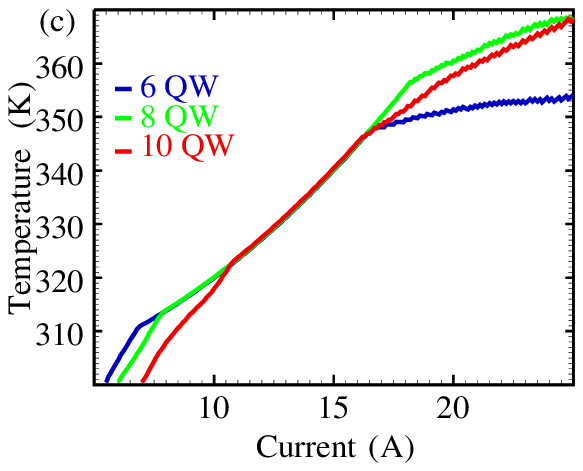}%
\includegraphics[width=0.5\textwidth]{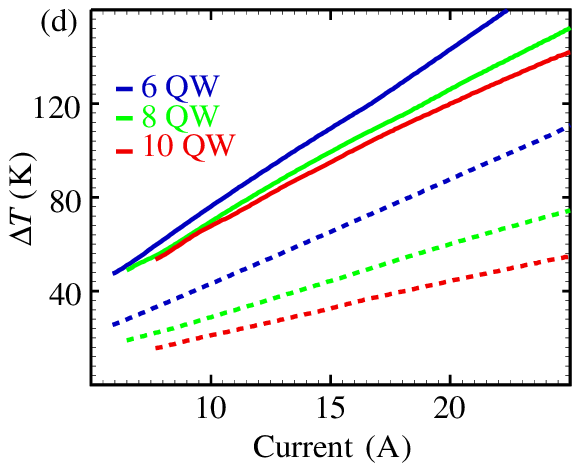}
\caption{
Calculated (a) Output power (b)  lasing wavelength, (c) average temperature of the active region at the end of 100~ns pump pulse and (d) electron heating in first and last 6, 8 and 10 (dotted) QW vs drive current for  resonator lengths $L=1$~mm. Resonator width is 100 $\mu$m, reflection coefficients of the facets are $r_1 =0.9$, $r_2 = 0.2$. The heat sink temperature is 300~K.
}\label{Fig_PI_HgCdTe}
\end{figure}

The output characteristics of the laser were calculated for pulsed mode operation with 100~ns pumping current impulse. During the impulse the active region of the structure is heated, so the momentary power and generation wavelength decrease to the end of the pulse Fig.~\ref{fig:P_t_I} \cite{2021/Afonenko/JPhysD_ApplPhys/Feasibility}. 
For a drive current of 16 A, the output power reduces from 1000 to 150 mW  during a 100 ns pulse. With a further increase in the drive current, the power drops to zero in a time shorter than the duration of the pump pulse.

Figure~\ref{Fig_PI_HgCdTe} shows the average power, wavelength and temperature of the active region as a function of pumping current for 1-mm long and 100-$\mu$m wide resonator. The threshold current increases from 6 to 8 A as the number of QWs increases from 6 to 10. The power 600 mW is reached at pumping current of approximately 20 A for a structure with 8 wells.

The temperature of the active region is determined by the injection current and practically does not differ for structures with different number of QWs. At a current of 20 A the average heating of the active region is $\sim$ 60 K. The range of the current-induced tuning of the generation wavelength is about 140 nm from $\sim$3.02 to $\sim$2.88 $\mu$m. This tunning is related to the temperature dependence of the bangap.

Non-equilibrium temperature of electrons and holes in the QWs was calculated using the phonon lifetime $\tauph=0.9$ ps which was estimated from experimental data on the phonon attenuation energy \cite{1974/Grynberg/PhysRevB/Dielectric,%
1991/Dingrong/JAP/Far-infrared,%
2011/Talwar/PRB/Infrared}. 
At Cd$_{0.85}$Hg$_{0.15}$Te barrier layers bandgap of 1.15 eV (300 K) the thermalization of each electron-hole pairs to the ground QW states (separated by 0.4 eV) is accompanied by the emission of more than 40 phonons with energy of $\sim 18$ meV. Calculations have shown that accounting for the effect of \lq\lq{}hot\rq\rq{} phonons leads to an increase in the non-equilibrium temperature of charge carriers in the QW by 40--140~K at 20~A depending on the number of QWs. The electronic temperature in different quantum wells may strongly differ because of different current injection into these wells (the injection coefficient drops with distance from p-emitter because of the low mobility of holes). The electron temperature decreases with the increase of the number of the QWs. 
Excess of electron temperature over lattice temperature about 150 K was observed experimentally \cite{Kudryavtsev_TempLimitations}.

\section{Conclusion} 

Thus, this paper proposes the design of an injection laser with a 2.2-nm thick HgTe QWs for generation in 3 $\mu$m region at room temperature. The analysis of structures with 6- to 10 QWs is based on a model that takes into account carrier drift and diffusion in barrier layers, capture in quantum wells, radiative and non-radiative recombination,  heating of the active region. Based on the developed technique, the refractive index of quantum-sized HgTe layers was calculated, which was 5.14 at a wavelength of 3 $\mu$m. The value of two-dimensional Auger recombination coefficient was calculated to be $4.7 \times 10^{-15}$~cm$^4$/s at lattice temperature 300~K. 
It is found that the contribution to the Auger recombination of transitions involving continuum states exceeds the contribution from transitions involving only localized states. The dominant mechanism of Auger recombination is a process involving two holes and one electron.
A model for calculating the non-equilibrium electronic temperature due to the emission of \lq\lq{}hot\rq\rq{} phonons was developed. It is found that at 0.9-ps optical phonon lifetime the temperature of electrons and holes in the QW can exceed the lattice temperature by 40--140~K depending on the number of QWs. It is shown that for structure with 8 QWs the maximum power of 600~mW is reached when pumped by 100 ns current pulses at pumping current of 20 A, the resonator length of 1 mm and the aperture of 100 $\mu$m. At the same time during the pumping pulse there is a shifting of generation wavelength of more than 120 nm from $\sim$3.02 to $\sim$2.88 $\mu$m because of active region lattice temperature increase of 60 K.

\section*{Acknowledgements}

This work was supported by the Ministry of Science and Higher Education of the Russian Federation (grant \# 075-15-2020-797 (13.1902.21.0024)).


\end{document}